\documentclass[conference]{IEEEtran}

\usepackage{times,epsfig,graphicx,wrapfig,bbm,color}
\usepackage{bm, amssymb, amsmath, amsfonts}

\usepackage{cite}
\def\BibTeX{{\rm B\kern-.05em{\sc i\kern-.025em b}\kern-.08em
    T\kern-.1667em\lower.7ex\hbox{E}\kern-.125emX}}
    
\usepackage{color}
\definecolor{red}{rgb}{1,0.2,0.2}
\definecolor{green}{rgb}{0.2,1,0.5}
\definecolor{blue}{rgb}{0,0,1}
\definecolor{lightblue}{rgb}{0.3,0.5,1}

\newcommand {\N} {{\rm I\kern-2.5pt N}}
\newcommand {\R} {{\rm I\kern-2.5pt R}}

\newtheorem{prop}{Proposition}

\newtheorem{theorem}{Theorem}
\newtheorem{assm}{Assumption}

\newcommand{\1}{\mathbf{1}}
\newcommand{\0}{\mathbf{0}}
\newcommand{\ab}{\mathbf{a}}

\newcommand{\eb}{\mathbf{e}}

\newcommand{\gb}{\mathbf{g}}

\newcommand{\yb}{\mathbf{y}}

\newcommand{\beqa}{\begin{eqnarray}}
\newcommand{\eeqa}{\end{eqnarray}}
\newcommand{\beqan}{\begin{eqnarray*}}
	\newcommand{\eeqan}{\end{eqnarray*}}
\newcommand{\beq}{\begin{equation}}
\newcommand{\eeq}{\end{equation}}

\newcommand{\bfl}{\begin{flushleft}}
	\newcommand{\efl}{\end{flushleft}}
\newcommand{\bgnv}{\begin{verbatim}}
	\newcommand{\ednv}{\end{verbatim}}

\newcommand{\myb}{\hspace{-0.1in}}
\newcommand{\myeq}{& \hspace{-0.1in} = & \hspace{-0.1in}}

\newcommand{\lb}{\nonumber \\}
\newcommand{\myarr}{\begin{array}{lll}}

	\newcommand{\cA}{{\cal A}}
	\newcommand{\cR}{{\cal R}}
	
	\newcommand{\cL}{{\cal L}}
	\newcommand{\cP}{{\cal P}}
	\newcommand{\cN}{{\cal N}}
	\newcommand{\cF}{{\cal F}}
	\newcommand{\cS}{{\cal S}}
	\newcommand{\be}{{\bf e}}
	\newcommand{\bx}{{\bf x}}
	\newcommand{\by}{{\bf y}}
	
	\newcommand{\bv}{{\bf v}}

\newcommand{\bg}{{\bf g}}

\newcommand{\bU}{{\bf U}}
\newcommand{\bV}{{\bf V}}

\newcommand{\bitem}{\begin{itemize}}
	\newcommand{\eitem}{\end{itemize}}
\newcommand{\benum}{\begin{enumerate}}
	\newcommand{\eenum}{\end{enumerate}}

\newcommand{\norm}[1]{\left| \left| #1 \right| \right|}

\newcommand{\bP}[1]{{\mathbbm P}\left[ #1 \right]}
\newcommand{\myhb}{\hspace{-0.25in}}

\newcommand{\avg}{{\rm avg}}

\begin{document}

\title{End-to-End Quality-of-Service Assurance with Autonomous Systems: 5G/6G Case Study}

\author{\IEEEauthorblockN{Van Sy Mai}
\IEEEauthorblockA{\textit{CTL, NIST} \\
Gaithersburg, MD USA\\
vansy.mai@nist.gov}
\and
\IEEEauthorblockN{Richard J. La}
\IEEEauthorblockA{\textit{ITL, NIST and} \\
\textit{ECE \& ISR, Univ. of MD}\\
Gaithersburg, MD USA \\
richard.la@nist.gov}
\and
\IEEEauthorblockN{Tao Zhang}
\IEEEauthorblockA{\textit{CTL, NIST} \\
Gaithersburg, MD USA \\
tao.zhang@nist.gov}
\and
\IEEEauthorblockN{Abdella Battou}
\IEEEauthorblockA{\textit{CTL, NIST} \\
Gaithersburg, MD USA \\
abdella.battou@nist.gov}
}

\maketitle

\begin{abstract}
Providing differentiated services to meet the unique requirements of different use cases is a major goal of the fifth generation (5G) telecommunication networks and will be even more critical for future 6G systems. Fulfilling this goal requires the ability to assure quality of service (QoS) end to end (E2E), which remains a challenge. A key factor that makes E2E QoS assurance difficult in a telecommunication system is that access networks (ANs) and core networks (CNs) manage their resources autonomously.  So far, few results have been available that can ensure E2E QoS over autonomously managed ANs and CNs. Existing techniques rely predominately on each subsystem to meet static local QoS budgets with no recourse in case any subsystem fails to meet its local budgets and, hence will have difficulty delivering E2E assurance. Moreover, most existing distributed optimization techniques that can be applied to assure E2E QoS over autonomous subsystems require the subsystems to exchange sensitive information such as their local decision variables. This paper presents a novel framework and a distributed algorithm that can enable ANs and CNs to autonomously ``cooperate" with each other to dynamically negotiate their local QoS budgets and to collectively meet E2E QoS goals by sharing only their estimates of the global constraint functions, without disclosing their local decision variables. We prove that this new distributed algorithm converges to an optimal solution almost surely, and also present numerical results  to  demonstrate that the convergence occurs quickly even with measurement noise.
\end{abstract}

\section{Introduction}

A major goal of the fifth generation (5G) telecommunication networks is to provide differentiated (or even optimized) services to meet different use cases' unique requirements, e.g., on performance, availability, and security \cite{TS22.261}. The ability to provide such differentiated support is a prerequisite for many IoT (Internet of Things) applications such as automated manufacturing, vehicle-to-vehicle communications, remote operation (teleoperation) of ground vehicles and drones, and telesurgery. It represents a fundamental advancement over the previous generations of networks that had focused on providing one-size-fits-all network connectivity for all applications.

To enable differentiated services, 5G shifted from a communication-centric architecture to a service-centric architecture that separated network and service control functions from the physical infrastructure, further modularized these functions, and defined their interactions in terms of providing and consuming services. This service-centric architecture enables network slicing that creates end-to-end (E2E) logical networks (network slices) dedicated for specific services, forming a foundation for delivering differentiated services. Current 5G specifications further describe a framework and its enabling protocols for collecting network performance data required to support service assurance \cite{TS28.550, TS28.552, 
TS28.554, TS32.425, TS28.532}. 

Despite recent advances, fundamental challenges remain to be addressed before the promise of differentiated services can be fully realized in 5G and future generations of telecommunication networks \cite{Saad20}. One such challenge is to assure E2E quality of service (QoS). Such assurance is essential to, for example, many delay-sensitive applications \cite{Zhang2020, Zhu19} and will be more critical for future 6G systems that will support more use cases and more stringent requirements \cite{Letaief19, Saad20}.

A key factor that makes E2E service assurance difficult in telecommunication networks is the fact that access networks (ANs) and core networks (CNs) manage their resources autonomously. While such autonomous management simplifies the tasks of designing, operating, and evolving a large system, limited coordination among ANs and CNs on resource management also leaves E2E service assurance more challenging~\cite{She20}. 

To achieve E2E QoS assurance over autonomously managed ANs and CNs, current 5G specifications decompose the E2E Key Performance Indicator (KPI) targets (e.g., packet delay, packet drop ratio) into KPI budgets for each CN and AN, and rely on each of them to autonomously manage its resources to satisfy its allocated KPI budgets. For example, 5G specifications decompose the E2E Packet Delay Budgets (PDBs) into CN PDBs and AN PDBs, with the CN PDBs for standardized QoS flows predefined in the specifications and to be statically configured in each CN \cite{TS23.501}. A network operator will then need to determine the AN PDBs for each AN, which is done statically today.

Static KPI budget allocations cannot effectively respond to dynamic network conditions or account for the varying characteristics of different networks (e.g., different network capabilities, capacities, and supported services). Also, since an AN or CN may violate its KPI budgets due to random events such as network congestion, network operators will need ways to ensure that such violations will not lead to intolerable violations of E2E KPI requirements. 

Recognizing the limitations of static KPI budget allocations, 5G permits non-standard QoS flows to support dynamic PDBs and allows the dynamic PDBs to be signaled between ANs and CNs. However, current 5G specifications do not prescribe how dynamic PDBs for each AN or CN should be determined or enforced, and how E2E PDB requirements can be assured in case any autonomous network along an E2E path violates its allocated PDB budgets. 

The past few years have also seen growing efforts in developing resource allocation mechanisms for assuring E2E QoS for 5G and beyond systems. Here, we summarize only the most closely related  studies  that  focused on E2E QoS rather than QoS for RANs or CNs alone. We emphasize that this is not meant to be an exhaustive list. In \cite{Lee19}, the 5G QoS architecture is extended to incorporate the Framework for Abstraction and Control of Traffic Engineered Networks (ACTN), which was developed by the IETF (IETF RFC 8453) to manage large multi-domain networks. This extended architecture calls for the resource controllers in different technology or administrative domains (e.g., ANs and CNs) to be orchestrated by a Multi-Domain Service Coordinator function to achieve E2E service assurance. But, it does not describe how such orchestration can be accomplished. In \cite{Teng20}, a high-level architecture for 5G service assurance is described. It focuses on monitoring E2E KPIs rather than providing mechanisms for achieving E2E service assurance. A dynamic resource slicing method aimed at E2E service assurance is proposed in \cite{Ye18b}, and it treats resource slicing in the radio ANs and the CN independently. A technique that combines user device QoS provisioning with network QoS provisioning is proposed in \cite{Asad20}. It treats QoS provisioning in ANs and CNs as independent tasks and relies on an unrealistic assumption that the KPI budgets for each network node are fixed and known. In \cite{Dutra17}, path configuration methods are proposed for ensuring E2E QoS in Software-Defined Networks (SDNs). SDNs are expected to be widely used in 5G systems. But, the path configuration methods in \cite{Dutra17} require centralized provisioning and control of E2E network paths, which is impractical in telecommunication systems where ANs and CNs are expected to continue to be managed separately.  

In order to assure E2E service quality while allowing the ANs and the CNs to control their resources independently, these autonomous networks should be able to coordinate with each other to meet E2E KPI requirements. To this end, we propose an optimization-based framework in which the E2E KPI requirements are captured by {\em global} constraints. Based on this framework, we design a new {\em distributed} algorithm; existing distributed optimization algorithms require the subsystems to maintain and exchange their local decision variables with each other. This can result in heavy signaling overheads, increase technology inter-dependency among ANs and CNs, or reveal how different subsystems manage their internal resources, and thus are impractical or undesirable in telecommunication systems \cite{Bianchi13, Jacquet2014, Srivastava11}. 

In contrast, our algorithm only needs the subsystems to exchange their {\em estimates of the global constraint functions}. When the subsystems update their local decision variables using the proposed algorithm, they converge to an optimal point with probability 1. We present numerical results to demonstrate that the proposed 
algorithm quickly reduces the cost of the system to
near the optimal value even with measurement noise. Finally, we illustrate that each optimal point gives rise to a stationary state Nash equilibrium of an associated state-based potential game.  

{\bf Notation:} We denote set of nonnegative integers 
by $\N$. Unless stated otherwise, all vectors are column vectors
and $\norm{\cdot}_2$ denotes the $\ell_2$ norm. 
Given a vector $\bx$, we denote the $k$th element 
by $x_k$. The vector of 
zeros (resp. ones) of appropriate dimension is 
denoted by $\0$ (resp. $\1$). 
Given a closed convex set ${\cal S}$ and a vector $\bx$, 
$P_{{\cal S}}(\bx)$ denotes the Euclidean projection of $\bx$ onto ${\cal S}$.

\section{Proposed Optimization Framework}
    \label{sec:Framework}

In the rest of the paper, we refer to
an AN or a CN as an agent, and let $\cA$ be the set of $N$ agents. The problem of ensuring E2E QoS over autonomously managed networks can be formulated as a distributed optimization problem with both {\em local} and {\em global} constraints. 
Let $\yb_i$ be the local decision variables of agent $i \in \cA$, which are used by agent $i$ to manage its network
resources. The local constraints of agent $i$ 
restrict $\yb_i$ independently of 
the decision variables chosen by other agents. We call the set of $\by_i$ that satisfies agent $i$'s local constraints its local constraint set and denote it by ${\cal G}_i$. Let $\by := (\by_i : i \in \cA)$ be the vector comprising the decision variables of all agents. The set of $\by$ that satisfies the local 
constraints of all agents is denoted by ${\cal G}
:= \prod_{i \in \cA} {\cal G}_i$. 

We capture the E2E QoS guarantees, which the agents 
must meet collectively, using $K$ global constraint functions $g_1, \ldots, g_K$. The global constraints depend on the 
decision variables of more than one agent and, hence, 
require their coordination. 

If (i) the local objective functions 
of individual agents, (ii) local constraint 
sets ${\cal G}_i$, $i \in \cA$, and 
(iii) global constraint functions $g_1, \ldots, 
g_K$ are known, we can formulate the problem of 
optimizing the overall network performance subject
to the E2E QoS requirements as a 
constrained optimization problem of the form given below, 
where (a) the objective function is separable, and (b)
global (inequality) constraints couple the decision
variables of agents:
\begin{subequations}    \label{eq:GameForm}
	\begin{eqnarray}\label{ProbCVX_Game}
    \min_{ \by \in {\cal G} } 
	&& \myb \sum_{i \in \cA} \phi_i(\by_i)  \\
	\mbox{s.t.} 
	&& \myb \gb(\yb) := \sum_{i \in \mathcal{A}} \gb_{i}(\yb_i) 
	    \le \0, \label{eqConstraint}
	\end{eqnarray}
\end{subequations}
where $\phi_i$ denotes the local objective function
that agent $i$ aims to minimize using its decision
variables $\yb_i$. Here, $\gb := (g_1, \ldots, g_K)$ 
is the vector function consisting
of $K$ global constraint functions. In our setting, 
each agent $i$ knows $\bg_i$, which represents 
its contributions to global constraint functions, 
but not $\gb_j$, $j \in \cA \setminus \{i\}$.

\subsection{Example: End-to-end Delay Constraints}
    \label{subsec:Example}

Consider a network consisting of $N$
domains or autonomous systems, which are the
agents in our framework, and 
	denote the set of agents by $\cA$.
	Suppose that the network provides $n_S$ different
	service classes and ${\cal S}$ is the set of 
	service classes. 
	Let ${\cal F}$ be the set of $n_F$ flows that 
	need to be supported. Each flow $f \in {\cal F}$
	is of service class $s_f \in {\cal S}$. Thus, 
	the set ${\cal F}$ can be partitioned into
	$\{ {\cal F}^s : s \in \cS \}$, where
	${\cal F}^s$ is the set of service class $s$
	flows in ${\cal F}$ with $n^s := |\cF^s|$, 
	$s \in \cS$. 
	Each flow $f$ traverses a subset of
	agents in $\cA_f$. We use an $n^s \times N$
	matrix ${\bf R}^s =[R^s_{f,a} : f \in \cF^s, 
	a \in \cA]$ to describe the routes taken
	by the flows of service class $s$, 
	where $R^s_{f,a} = 1$ if flow $f$
	traverses agent $a$, and $R^s_{f,a} = 0$ otherwise.

	\paragraph{Problem for individual agents} 
	Let $\cN^a$ be the set of egress/ingress nodes
	in agent $a$ and $m^a := |\cN^a|$. 
	An agent $a$'s flow $b$ refers to a triple $(s, 
	p_i, p_e) \in \cS \times \cN^a \times \cN^a$
	with the understanding that it carries
	service class $s$ traffic that enters and
	leaves agent $a$ at $p_i$ (ingress point) 
	and $p_e$ (egress point), respectively. 
	We denote the set of flows of service class
	$s$ in agent $a$ by 
	$\tilde{\cF}^a_s$, and let $\tilde{\cF}^a := \cup_{s \
	\in \cA} \tilde{\cF}^a_s$. 
	
	For each agent $a \in \cA$, there is a fixed
	traffic tensor that describes the amount
	of traffic that needs to be transported
	for each ingress-egress pair and each service
	class. We denote this traffic tensor by ${\bf T}^a$, 
	which is an $n_s \times m^a \times m^a$ tensor; 
	the element $T^a_{s, p_i, p_e}$ denotes the demand
	of flow $b = (s, p_i, p_e)$, i.e., it is the amount 
	of traffic belonging to service class $s \in \cS$, 
	which needs to be transported from ingress node 
	$p_i \in \cN^a$ to egress node $p_e \in \cN^a$.
	We use $I(b)$ and $E(b)$ to denote the ingress
	and egress node, respectively, of flow $b$. 
	
	Let $\cL^a$ be the set of (directed) 
	links in agent $a$. Each link $\ell 
	\in \cL^a$ has a finite capacity 
	$c^a_\ell$, and ${\bf c}^a := (c^a_\ell
	: \ell \in \cL^a)$. 
	Each flow $b \in \tilde{\cF}^a$ is supported by a
	set of routes in $\cR^a_b$ connecting 
	$I(b)$ to $E(b)$, and 
	a route $r \in \cR^a_b$
	utilizes a subset 
	of links in agent $a$, which we denote
	by $\cL^a(r)$ with a little abuse of
	notation.\footnote{Note that two routes
		used for two distinct flows may utilize
		the same set of links. In this case, 
		we treat them as
		two separate routes.} 
	For each $s \in \cS$, 
	we define $\cP^a_s := 
	\cup_{b \in \tilde{\cF}^a_s} \cR^a_b$ and a
	routing matrix ${\bf R}^a_s$, which is a 
	$|\cP^a_s|
	\times |\cL^a|$ matrix; the element 
	$R^a_{r, \ell} = 1$ if $\ell \in \cL^a(r)$
	and $R^a_{r, \ell} = 0$ otherwise.
	
	The variable $x^a_{b, r}, r \in \cR^a_b$
	for some flow $b \in \tilde{\cF}^a$, denotes the
	amount of flow $b$ traffic traversing 
	route $r$. Define $\bx^a = (x^a_{b, r} : b \in 
	\tilde{\cF}^a \mbox{ and } r \in \cR^a_b)$ to
	be the {\em local} decision variables of agent $a$.
	These local decision variables must be 
	nonnegative and support
	the given traffic tensor: for every 
	$b \in {\cal F}^a$,
	$$
    \textstyle x^a_{b, r} \geq 0
    \mbox{ for all } r \in {\cal R}^a_b \ \mbox{ and }  
    \ \sum_{r \in {\cal R}^a_b} x^a_{b,r} 
	= T^{a}_b.
	$$ 
	We use ${\cal G}_a$ to denote the set of agent $a$'s
	decision variables that satisfy these conditions.

	The goal of agent $a$ is to minimize its own 
	{\em local} cost given by $C^a(\bx^a)$.
	This local cost $C^a(\bx^a)$ can be, for
	example, the (weighted)
	average delay experienced by the traffic
	traversing agent $a$ or the costs associated with packet
	losses suffered in its network. Oftentimes, this
	cost includes a summable cost, i.e., the sum of the 
	costs of individual links, which can be written as 
	$\textstyle
	\sum_{s \in \cS} \left( 
	\textstyle \sum_{\ell \in {\cal L}^a}
	c_{s, \ell}(\bx^a)
	\right)$,
	where $c_{s,\ell}(\bx^a)$ is the cost of class 
	$s$ traffic on link $\ell$ as a function of 
	$\bx^a$.

	\paragraph{E2E delay requirements}
	The traffic of service class $s$ cannot
	tolerate (average) E2E delays larger than $d^*(s)$.
	We handle these E2E delay constraints using 
	the following {\em global} constraints: suppose that, 
	given local decision variables $\bx^a$ of 
	agent $a$, the 
	delay experienced by service class $s$ traffic on 
	link $\ell$ is given by $D^{a}_{s, \ell}(\bx^a)$. 
	Then, the delays experienced by flows of service 
	class $s$ through agent $a$ are given by	
	\beqan
	{\bf R}^a_{s} {\bf D}^a_s(\bx^a), 
	\eeqan
	where ${\bf D}^a_s(\bx^a) := (D^a_{s, \ell}(\bx^a) 
	: \ell \in {\cal L}^a)$. In general, it is
	reasonable to assume that $D^a_{s, \ell}(\bx^a)$
	is a convex function of $\bx^a$. We define 
$$
g_{a,s}(\bx^a) := \max({\bf R}^a_{s} {\bf D}^a_s(\bx^a)), 
\quad s \in {\cal S},
$$ 
to be the maximum delay experienced by service class $s$
traffic in agent $a$'s network.

Let $\bx := 
(\bx^a : a \in \cA)$ and 
${\cal G} := 
	    \prod_{a \in \cA} {\cal G}_a$. 
The problem of minimizing the total cost of all
agents can be set up as the following 
constrained optimization problem: 
\begin{subequations}    \label{eq:CombinedProb2}
\beqa
\myhb \min_{ \bx \in {\cal G} } 
	    && \myb \sum_{a \in \cA} C^a(\bx^a)
	    \label{eq:AS_obj2}   \\
\myhb \mbox{s.t.} && \myb \sum_{a \in \mathcal{A}} R^s_{f,a} g_{a,s}(\bx^a) \leq d^*(s), 
	\ s \in \cS \mbox{ and } f\in \mathcal{F}^s
	    \label{eq:Delay_constr2}
\eeqa
\end{subequations}
It is evident that the above problem in \eqref{eq:CombinedProb2} 
is of the form in \eqref{eq:GameForm}. Note that the decisions 
by the agents are coupled only via the constraints in 
\eqref{eq:Delay_constr2}, while the objective function is 
separable.

\section{Distributed Optimization Algorithm with Noisy
    Measurements}
    \label{sec:Convergence}
    
It is clear from the discussion and the example in the previous section that, in order for the agents to optimize the overall network performance with global constraints, we need a new distributed optimization framework that will allow the agents to cooperate with each other to satisfy the global constraints without incurring prohibitive overheads for information exchange.
Furthermore, in order to allow different agents to use different networking technologies, equipment, and equipment vendors, it will be important to eliminate or minimize the need for the agents to exchange sensitive information that may reveal how they manage their networks.

\subsection{Penalty Method with Noisy Measurements}

One popular approach to solving a constrained optimization problem is a penalty method, which adds a penalty function to 
the objective function. The penalty increases
with the level of violations of constraints. 
Although a general penalty 
function can be used in our problem, we assume a 
specific penalty function to simplify our exposition:
consider the following approximated problem 
with a penalty function:
\begin{equation}    \label{eq:ApproximatedProblem}
\min_{ \by \in {\cal G} } \quad 
	\sum_{i \in \cA} \phi_i(\by_i)  
	    + \frac{\mu}{2N} \sum_{k \in {\cal K}_G}
	        \big( [g_k(\yb)]_+ \big)^2
    =: \Phi(\yb),
\end{equation}
where $\mu>0$ is a penalty parameter,
${\cal K}_G$ is the set of $K$ global constraints, 
$\phi_i$ is the (local) objective function of
agent $i$ (introduced in \eqref{eq:GameForm}), and 
$[\cdot]_+ := \max(0, \cdot)$. 
The original 
problem in \eqref{eq:GameForm} is recovered by 
letting $\mu \to \infty$. The function $\Phi$ is 
convex on ${\cal G}$ provided that $\phi_i$ and 
$g_{i,k}$, $i \in \cA$ and $k \in {\cal K}_G$, are convex.

Unfortunately, the usual gradient projection method 
does not lead to a distributed algorithm for our
problem because the 
penalty function and its gradient are coupled. 
Also, in many problems of practical interest,
(a) the analytic expression of $\phi_i$
and its gradient are unknown to agent $i$ and
(b) only noisy measurements are available to 
estimate them. 
For instance, in the example discussed in 
the previous section, an agent needs to estimate
its local costs from network measurements 
(e.g., packet loss rates), which contain 
measurement noise. 

In order to deal with the issues of (i) noisy estimates
of local costs and (ii) global constraints that couple
the decision variables of agents, we propose a new 
distributed optimization algorithm that requires each 
agent $i$ to keep an estimate $\eb_i(t)$ of the 
constraint functions $\bg(\yb(t))$. 
These estimates are used to approximate the 
gradient of the penalty function and are updated
using a consensus-type algorithm. 

\paragraph{Update rule for decision variables}
Consider the following iterative update rule, which is 
used by each agent to update its local decision variables: 
for every agent $i \in {\cal A}$ and $t \in \N$, 
\beqa   \label{eq:update_y}
\yb_{i}(t+1)
\myeq P_{{\cal G}_i} \big( \yb_{i}(t) 
    + \gamma_t \bU_{i}(t) \big), 
\eeqa
where $\bU_{i}(t)$ are random vectors, and
$\gamma_t$ is a step size at iteration $t$. 
The initial value $\yb_i(0)$ is chosen arbitrarily 
in $\mathcal{G}_{i}$. 
The random vector can be decomposed as follows:
\beqa
&& \hspace{-0.3in} \bU_{i}(t) 
= - \nabla \phi_i(\yb_{i}(t))
    \! - \! \frac{\mu}{N} \sum_{k \in {\cal K}_G}
    \!\!\nabla g_{i,k}(\yb_i(t)) 
        [g_k(\yb(t))]_+ 
    \label{eq:U} \\
&& \myb \myb + \mu \sum_{k \in {\cal K}_G}
    \!\!\nabla g_{i,k}(\yb_i(t)) 
    \Big( \frac{1}{N} [g_k(\yb(t))]_+ 
        \! - \! [e_{i,k}(t)]_+ \Big) 
        \! + \!  {\bf V}_{i}(t) \lb
&& \hspace{-0.3in}
= - \nabla \phi_i(\yb_{i}(t))
\! - \! \mu \sum_{k \in {\cal K}_G}
    \!\!\nabla g_{i,k}(\yb_i(t)) [e_{i,k}(t)]_+
\! + \!  {\bf V}_{i}(t). 
    \nonumber
\eeqa
The first and second terms on the right-hand side (RHS)
in \eqref{eq:U} are the gradient of the local objective function 
$\phi_i$ and the penalty function associated with 
global constraints, respectively. 
The third term models the 
difference between the correct gradient of
the penalty function and the estimated gradient
on the basis of agent $i$'s estimates of constraint 
functions, namely $\eb_i(t)$. The last term 
represents noise or stochastic perturbation in the 
estimates ${\bf U}(t) := (\bU_i(t): i \in \cA)$, 
the statistical distribution of 
which may depend on the current decision 
variables $\by(t)$. 
We emphasize that the exact gradients in \eqref{eq:U} 
are unavailable to agent $i$ and only a noisy estimate 
given by $\bU_i(t)$ is available, which need to be
computed from measurements. 

\paragraph{Update rule for global constraint function 
estimates} In addition to updating local decision variables
according to \eqref{eq:update_y}, each agent also
updates its estimate of global constraint functions using
the following update rule:
\beqa
\myhb \eb_i(t+1)
\myeq \sum_{j \in \mathcal{A}} w_{ij} \eb_j(t) 
    + \gb_i(\yb_i(t+1)) - \gb_i(\yb_i(t)),
    \label{eq:update_e}
\eeqa
where $W = [w_{ij} : i, j \in \cA]$ is the 
weight matrix, and 
$\eb_i(0) = \gb_i(\yb_i(0))$, $i \in \cA$. 

Clearly, the first term on the RHS of 
the consensus-type algorithm in \eqref{eq:update_e} 
requires exchanging the estimates of constraint 
functions maintained by agents, but not their local
decision variables. Furthermore, 
each agent $i\in \mathcal{A}$ 
needs to exchange its estimate $\eb_i$ only with its 
neighbors specified by the weight matrix $W$. 
This update rule is designed to track the 
average of the constraint functions. Moreover, under the 
assumption that the weight matrix $W$ is 
doubly-stochastic, we have
\begin{eqnarray} 
\eb_\avg(t) 
:= \frac{1}{N} \sum_{i \in \mathcal{A}} \eb_i(t) 
\myeq \frac{1}{N} \sum_{i \in \mathcal{A}} \gb_i(\yb_i(t)) 
    \lb
\myeq \frac{1}{N} \gb(\yb(t))
=: \bar{\gb}(t).
    \label{eqAverageError}
\end{eqnarray} 
As a result, the sum of the estimates maintained by all 
agents equals the correct value of constraint functions, 
and we can view $\eb_i(t)$ as a local estimate 
of the average constraint functions. We note that, 
since agents exchange their estimates only with 
(direct) neighbors, the communication overheads 
scale linearly with the number of global constraints.

In \eqref{eq:update_e}, we assumed that the same weight matrix $W$ is used for all constraints to simplify our discussion. In practice, however, agents may contribute to different sets of constraints and may not want to keep an estimate of constraint functions to which they do not contribute. In this case, we can adopt different weight matrices, one for each set of constraints with the same contributing agents.


\subsection{Convergence Results}    
    \label{subsec:Convergence}

In this subsection, we introduce the assumptions under
which agents' decision variables converge almost surely 
to an optimal point of \eqref{eq:ApproximatedProblem} 
when they adopt the proposed algorithm given by 
\eqref{eq:update_y} and \eqref{eq:update_e}. 

\begin{assm}    \label{assm:1}
(a) The local constraint sets ${\cal G}_i$, 
$i \in \cA$, are nonempty, compact and convex; 
and
(b) the local objective functions $\phi_i$ and 
constraint functions $g_{i,k}$, 
$i \in \cA$ and $k \in {\cal K}_G$, are convex 
and Lipschitz continuous on ${\cal G}_i$.
\end{assm}

Assumption~\ref{assm:1} is likely to hold in 
many, if not most, networking-related problems, such 
as resource allocation and performance optimization 
problems.

\begin{assm}    \label{assm:2}
The directed graph associated with 
the weight matrix $W$ is strongly connected, and the 
weight matrix $W$ is doubly stochastic.
\end{assm}

This assumption simply asserts that the agents can communicate with each other over one or more hops. When this assumption is violated, it would not be possible for some agents to cooperate with each other, making it 
difficult to satisfy global constraints in dynamic
environments.

\begin{assm}    \label{assm:3}
Given $\yb(t) = \by$, the distribution 
of the perturbation $\bV(t)$ is $\mu_{\yb}$ and satisfies 
(a) $\int \bv \ \mu_{\yb}(d\bv)$ 
$= \0$ for every $\yb \in {\cal G}$ and (b)
$\sup_{\yb \in {\cal G}} \int \norm{ {\bf v} }^2 
    \mu_{\yb}(d{\bf v}) =: \nu < \infty$.   
\end{assm}

This assumption states that, although measurements are
noisy, (a) if the measurements are taken many times, 
the average value will be close to the correct value
and (b) the noise is not large enough to prevent the
agents from extracting information pertinent to the
network performance. For example, when the noise is 
bounded, the second part of the assumption holds, which 
is likely the case in practice.  

{\bf Example:} Before we proceed, we explain what 
these assumptions mean in the context of the example
discussed in Section~\ref{subsec:Example}. 
The local problems described in Section
\ref{subsec:Example} are
essentially a (probabilistic) routing problem. 
Therefore, for each ingress-egress pair 
(with positive traffic), the feasible local 
decision variables lie in an appropriate 
standard simplex, 
which is compact and convex. Assumption
\ref{assm:2} means that the network is 
connected so that traffic can be routed 
from one agent to any other.
It is reasonable to assume that network
measurements are unbiased, i.e., the sample average
will converge to the correct value when the 
number of samples is large. In this case, 
Assumption~\ref{assm:3}a holds. 
Further, in practice the network 
measurements are bounded and, consequently, 
Assumption~\ref{assm:3}b will be true. 


Let us denote the optimal set, i.e., the set of optimal
points of \eqref{eq:ApproximatedProblem}, and the
optimal value by ${\cal Y}^*$ and $\Phi^*$, respectively. 
The following result asserts that, 
under Assumptions 1 through 3, the decision
variables updated in a distributed manner by the
agents converge to an optimal point almost surely. 
Its proof can be found in 
\cite{Mai2021}. 
Define $d_{{\cal Y}^*}:{\cal G} \to \R_+$, where 
$$
d_{{\cal Y}^*}(\by) 
= \inf\{ \norm{\by - \by^*}_2 \ | \ \by^* \in {\cal Y}^*\}.
$$

\begin{theorem} \label{thm:ASC1}
Suppose that Assumptions~\ref{assm:1} through 
\ref{assm:3} hold.
Under our proposed algorithm given by 
\eqref{eq:update_y} and \eqref{eq:update_e} 
with step sizes $\gamma_t$ satisfying  
$\sum_{t \in \N} \gamma_t \!=\! \infty$ and 
$\sum_{t \in \N} \gamma_t^2 \!<\! \infty$,
we have 
\begin{eqnarray*}
\bP{ 
\textstyle \lim_{t \to \infty} d_{{\cal Y}^*}(\yb(t))
= 0 } = 1.
\end{eqnarray*}
\end{theorem}

Note that the step size condition in this theorem 
is standard for many 
iterative (stochastic) optimization algorithms
and is needed to ensure the almost sure convergence
under the proposed algorithm.
In practice, however, a fixed step size may 
be used instead, and it has been shown 
in many cases that weak convergence hold, 
i.e., the optimization variables converge
to a small neighborhood around the optimal
set and remains there with high probability, 
when a sufficiently small constant step 
size is adopted~\cite{KushYin}. 



\section{Numerical Studies}
    \label{sec:Numerical}

In this section, we provide numerical studies 
to illustrate how our proposed framework and algorithm described in Sections~\ref{sec:Framework} and 
\ref{sec:Convergence} can be used to assure E2E QoS over a telecommunication system with autonomously managed ANs and CNs. This simplified scenario highlights the key benefits of the proposed framework and some practical considerations in applying such distributed optimization methods to real world problems. 

\begin{figure}[h]
\centerline{
\includegraphics[width=3.5in]{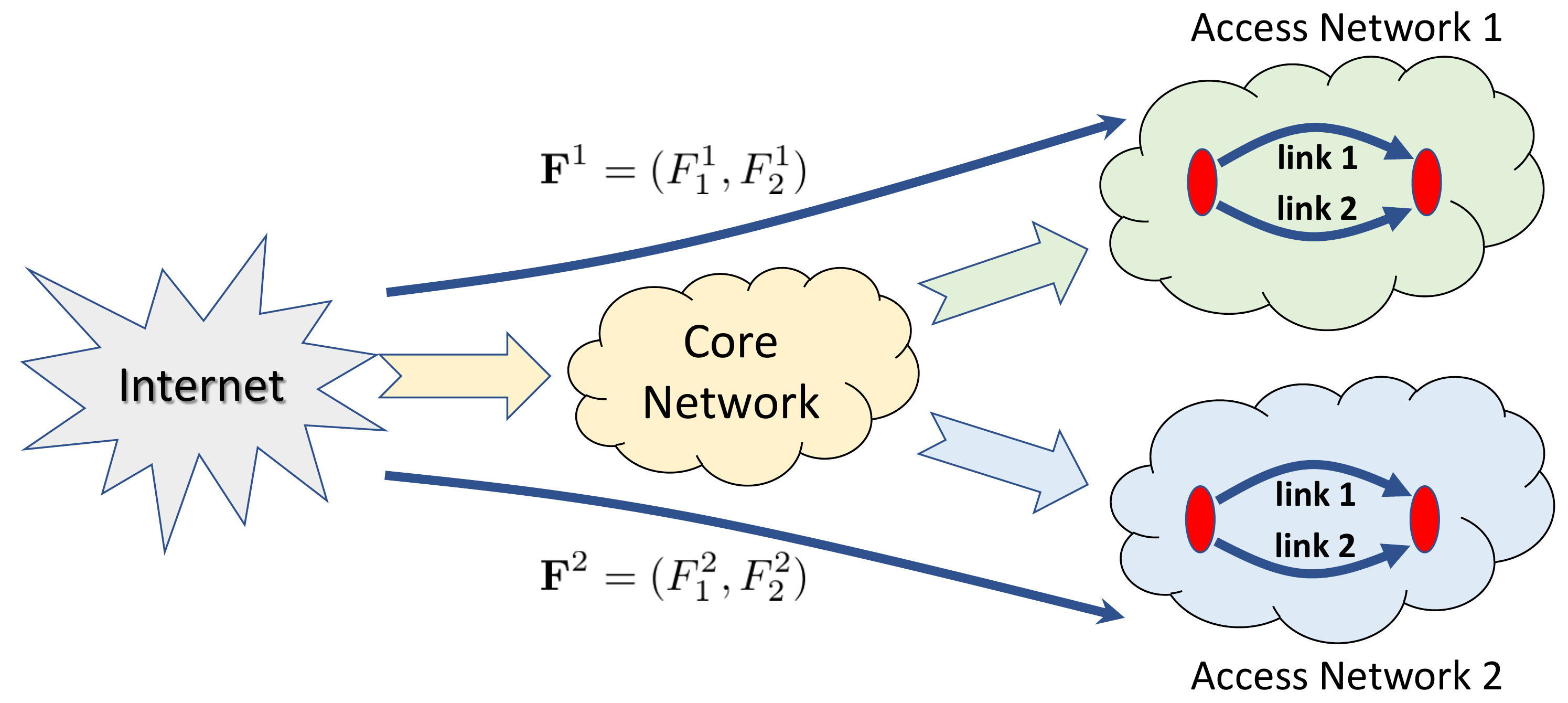}
}
\caption{Network used for numerical studies.}
\label{fig:example}
\end{figure}

\subsection{Description of setup}

We consider one CN and two ANs - AN 1
and AN 2, as shown 
in Fig.~\ref{fig:example}. For the numerical 
study, we only consider downstream traffic that enters 
the CN from the Internet and is destined
for either AN 1 or AN 2. The networks support two traffic classes, which require 
different average E2E delay guarantees.
The traffic destined for AN $i$ is given by 
${\bf F}^i = (F^i_1, F^i_2)$, where $F^i_s$ is
the class $s$ traffic destined for AN $i$.

Each AN has two links to carry traffic (links
1 and 2 shown in Fig.~\ref{fig:example}). The 
local decision variables of AN $i$ are 
$\by_i = \big( (r^i_{\ell, 1}, 
r^i_{\ell, 2}, b^i_\ell) : \ell = 1, 2\big)$, where 
$r^i_{\ell, s}$ is the fraction of class $s$ traffic
routed through link $\ell$ by AN $i$, and 
$b^i_\ell$ is the amount of bandwidth provisioned
on link $\ell$ by AN $i$. 
The cost of AN $i$ comprises (i) the cost of 
reserving $b^i_\ell$, $\ell = 1, 2$, on each 
link $\ell$ and (ii) the cost that is dependent on the average 
delays experienced by its traffic. 
More precisely, the cost of 
AN $i$, $i = 1, 2$, is given by 
\beqa
\phi_i(\by_i)
= \sum_{\ell = 1}^2 \Big( p^i_\ell(b^i_\ell)
    + \sum_{s=1}^2 \xi^i_s(r^i_{\ell,s} F^i_s, 
    d^i_\ell(f^i_\ell, b^i_\ell)) \Big), 
    \label{eq:ANcost}
\eeqa
where $f^i_\ell = r^i_{\ell, 1} F^i_1 + r^i_{\ell, 2} 
F^i_2$ is the total flow rate on link $\ell$, $p^i_\ell$
is the function that determines the total
reservation cost on link $\ell$,
$d^i_\ell$ determines the average delay on 
link $\ell$ as a function of reserved bandwidth
and the total flow rate, and $\xi^i_s$ is the delay cost function of class $s$ traffic.

Similarly, the local decision variables of the CN are the amounts of the
bandwidth provisioned for each traffic class and are given by $\by_c = (b_1, b_2)$, where $b_s$
is the amount of bandwidth reserved for class $s$ traffic. For fixed $\by_c$ and traffic ${\bf F}
= (F_1, F_2)$, where $F_s = F^1_s + F^2_s$ 
is the total amount of class $s$ traffic handled
by CN, its cost is given by 
\beqa
\phi_{c}(\by_{c}) 
\myeq \textstyle \sum_{s = 1}^{2} \big( p_s^{c}(b_s)
    + \xi^{c}_s(F_s, d^{c}_s(F_s, b_s)) \big), 
    \label{eq:CNcost}
\eeqa
where $p_s^{c}(b_s)$ is the cost of provisioning
bandwidth $b_s$ for class $s$ traffic,
$d^{c}_s$ determines the delay as a function of the total rate and provisioned
bandwidth for class $s$ traffic, and $\xi^c_s$ is the delay
cost function for class $s$ traffic.

The CN and the ANs collectively must satisfy constraints on the
average E2E delays experienced by the traffic. 
Specifically, the average E2E delay for traffic class $s$ should not exceed $D_s$. This gives
rise to the following {\em global} constraints:
\beqa
g_{i,s}(\yb_i, \yb_{c})
& \hspace{-0.1in} := & \hspace{-0.1in} 
\frac{1}{F^i_s} \sum_{\ell=1}^2 f^i_{\ell,s} 
    d^i_\ell(f^i_\ell, b^i_\ell)
+ d_s^{c}(F_s, b_s) - D_s
\leq 0, \lb
&& \hspace{0.5in} i = 1, 2 \ \mbox{ and } \ s = 1, 2, 
    \label{eq:constr1}
\eeqa
where $f^i_{\ell, s} = r^i_{\ell, s} F^i_s$ is
the amount of class $s$ traffic traversing link $\ell$. 

The approximated optimization problem we are interested
in solving is given by 
\beqa
\min_{\yb \in {\cal G}} 
    \Big( \big( \phi_c(\by_c) + \sum_{i=1}^2 \phi_i(\by_i) \big)
    + \frac{\mu}{6} \sum_{i=1}^{2} \sum_{s=1}^2
        \big( [ g_{i,s}(\by) ]_+ \big)^2 \Big)
    \label{eq:example_Phi}
\eeqa
with decision variables $\by = (\by_c, \by_1, 
\by_2)$.

\subsection{Practical Modifications of the Proposed
    Algorithm}
    \label{subsec:Modifications}

Our proposed framework and convergence
results in Sections~\ref{sec:Framework}
and \ref{sec:Convergence}, respectively, guarantee
the almost sure convergence of the decision variables
to an optimal point when the approximated problem
in \eqref{eq:ApproximatedProblem} is convex.
However, there are 
a few practical issues that one should consider, 
which are discussed here. 

\noindent $\bullet$ {\bf Modification 1 --} 
While the step sizes $\gamma_t$, $t \in \N$, 
can be chosen sufficiently small to avoid large
fluctuations in the decision variables, which are
updated in accordance with \eqref{eq:update_y} and 
\eqref{eq:update_e}, selecting small step sizes
also hampers their convergence to the optimal set. 
In order to skirt this issue, we apply a limiter
to (some elements of) $\gamma_t \bU_{i}(t)$ in 
\eqref{eq:update_y}. Specifically, we use the 
limiter to the elements corresponding to 
the routing probabilities $r^i_{\ell, s}$ so 
that the elements of $\gamma_t \bU_i(t)$ 
corresponding to these decision variables
always lie in [-0.01, 0.01]. The use of such
limiters alters our proposed algorithm from 
that described in Section~\ref{sec:Framework}. 
However, the almost sure convergence to the
optimal set shown in Section~\ref{sec:Convergence}
still holds with the modified algorithm
under the assumed Lipschitz continuity of 
gradients functions and bounded noise. 

\noindent $\bullet$ {\bf Modification 2 --} 
We observe from \eqref{eq:U} that
when agent $i$'s estimate of the $k$th 
global constraint at iteration $t$, namely 
$e_{i,k}(t)$, is positive, 
it introduces the corresponding term  
$\mu \nabla g_{i,k}(\by(t)) [e_{i,k}(t)]_+$ 
in the update term used in \eqref{eq:update_y}. 
Hence, when $e_{i,k}(t)$
oscillates around zero, it 
is added to the update term whenever $e_{i,k}(t) > 0$
while it makes no contribution to the update term
when $e_{i,k}(t) \leq 0$, causing a non-negligible
change in the update term. This is undesirable
and slows down the convergence of decision
variables. 

In order to cope with this issue, rather than 
using the true delay budgets $D_s, s = 1, 2$, we
use {\em target} delay budgets equal to 
$D^T_s = \tau D_s$, where $\tau \in (0, 1]$.
We call the global constraints with the delay
budgets replaced by the target delay budgets
the {\em fictitious} global constraints. 
The main idea behind the use of fictitious
global constraints is two-fold. First, satisfying the global constraints with true delay budgets in general requires selecting a large penalty parameter, which increases the sensitivity to the constraint function values (and their estimates). Second, using 
{\em fictitious} constraint functions allows
us to use a smaller penalty parameter (hence, 
reduces the aforementioned sensitivity to the global constraint
function values); we can violate the fictitious delay 
constraints while still satisfying the true delay 
constraints. This implies that the agents' estimates of
fictitious constraint functions stay
positive, avoiding the aforementioned oscillations,
while still satisfying the original global
constraints.

\subsection{Numerical Results}
    \label{subsec:NumericalResults}

In our numerical study, we assumed following functions:
\beqan
& \myb d^i_\ell(f, b) = \upsilon_\ell \big( f / b \big)^a, \ 
d_s^{c}(f, b) = \big( f / b \big)^a,
p^i_\ell(b) = \kappa^i_\ell \cdot b^k, & \\
& \myb p_s^{c}(b) = \kappa_s \cdot b^k, \ 
\xi^{c}_s(f, d) = \beta_s  f d, \ 
\mbox{and} \  
\xi^i_s(f, d) = \beta^i_s f d, &
\eeqan
where $\kappa^i_\ell > 0$, $\kappa_s > 0$, 
$\gamma_s > 0$, $\beta^i_s > 0$, 
$a \geq1 $ and $k \geq 1$.  
The following parameter values are used in the study:
\beqan
& \upsilon_1 = 1, \ \upsilon_2 = 0.8, \ 
\kappa^1_1 = 4, \ \kappa^1_2 = 5, \ 
\kappa^2_1 = 7, \ \kappa^2_2 = 9, & \\ 
& \kappa_1 = 4, \ \kappa_2 = 6,  
a = 2.5, \ k = 1.1, \ 
\beta^i_1 = \beta_1 = 40, & \\
& \mbox{ and } \ \beta^i_2 = \beta_2 = 10, \
i = 1, 2. & 
\eeqan

The flow rates are ${\bf F}^1 = (40, 20)$
and ${\bf F}^2 = (50, 30)$.
The delay budgets are chosen to be
$D_1 = 0.7$ and $D_2 = 0.5$ with the target
delay budgets set to 60 percent of the true
delay budgets, which are $D^T = (0.42, 0.3)$. 
The penalty parameter is set to $2 \cdot 10^4$:
a relatively large value of penalty parameter
is necessary in our scenario because the
delay budgets are much smaller than the costs
of reserving bandwidth for the networks. 
The employed doubly-stochastic weight matrix $W :=[ w_{ij} : 
i, j \in \{c, 1 ,2\}]$ used in 
\eqref{eq:update_e} is given by 
\beqan
W
= \begin{bmatrix}
3/4 & 1/8 & 1/8 \\
1/8 & 7/8 & 0 \\
1/8 & 0 & 7/8
\end{bmatrix} .
\eeqan

The stochastic perturbation ${\bf V}_i(t)$ introduced in the
noisy observation of gradient is uniformly
distributed over $[- \sigma \nabla \phi_i(\yb_i(t)), 
\sigma \nabla \phi_i(\yb_i(t))]$. For the reported numerical
results, we pick $\sigma = 0.75$. The 
step sizes are selected to be $\gamma_t = 
\min(0.1, (t+1)^{-0.6})$, 
$t \in \N$. Finally, the decision
variables are initialized as follows. Let
$F^i = \sum_{s=1}^2 F^i_s$, $i = 1, 2$, and 
$F^T = F_1 + F_2 = F^1 + F^2$:
\beqan
& \yb_c = (B_1, B_2), \ 
\yb_1 = (0.5, 0.5, 0.5, 0.5, B^1_1, B^1_2), & \\
& \mbox{and } \ 
\yb_2 = (0.5, 0.5, 0.5, 0.5, B^2_1, B^2_2), & 
\eeqan
where $B_s \sim$ Uniform($0, 2 F^T$), $s = 1, 2$, 
and $B^i_\ell \sim$ Uniform($F^i, 2 F^i$), $i = 1, 2$
and $\ell = 1, 2$. These random variables are 
selected independently. 

\begin{figure}[h]
\centerline{
\includegraphics[width=3.35in]{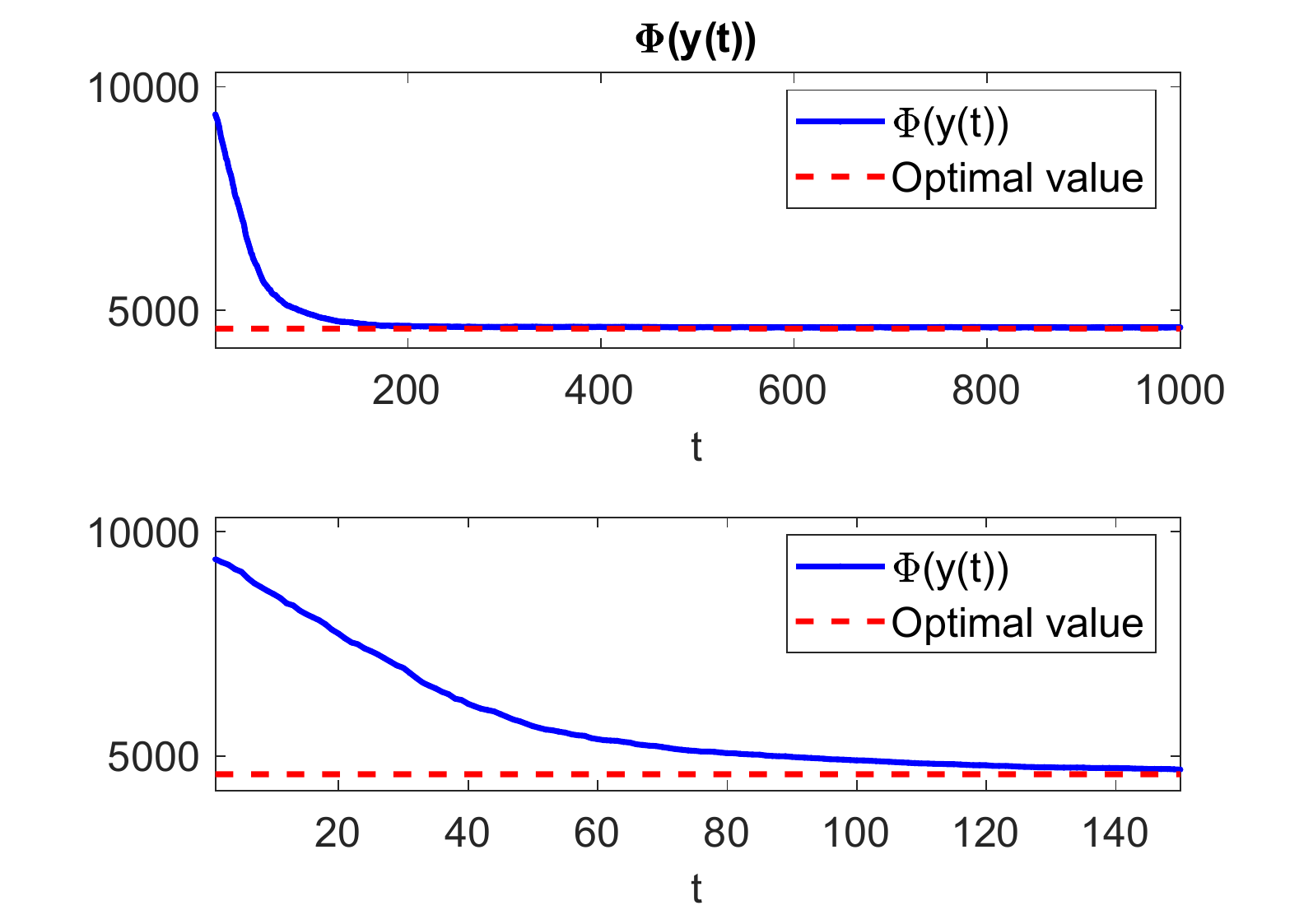}
}
\caption{Plot of the objective function 
$\Phi(\yb(t))$.}
\label{fig:Cost}
\end{figure}

First, we plot the value of the objective
function $\Phi(\yb(t))$ of the approximated
optimization problem in \eqref{eq:example_Phi} 
for $t$ in \{1, 2, \ldots, 1000\}
(top) and \{1, 2, \ldots, 150\} (bottom)
in Fig.~\ref{fig:Cost}. The dotted red lines
represent the optimal value we obtained using
the MATLAB optimization tool.\footnote{Mention of commercial products in this paper is for information only; it does not imply recommendation or endorsement by NIST.}
The plots suggest that
the value of $\Phi(\yb(t))$ converges to the
optimal value quickly. In fact, the bottom plot 
indicates that we are within 2-3 percent of the
optimal value within 150 iterations,
despite large noises. 

\begin{figure}[t]
\centerline{
\includegraphics[width=2in]{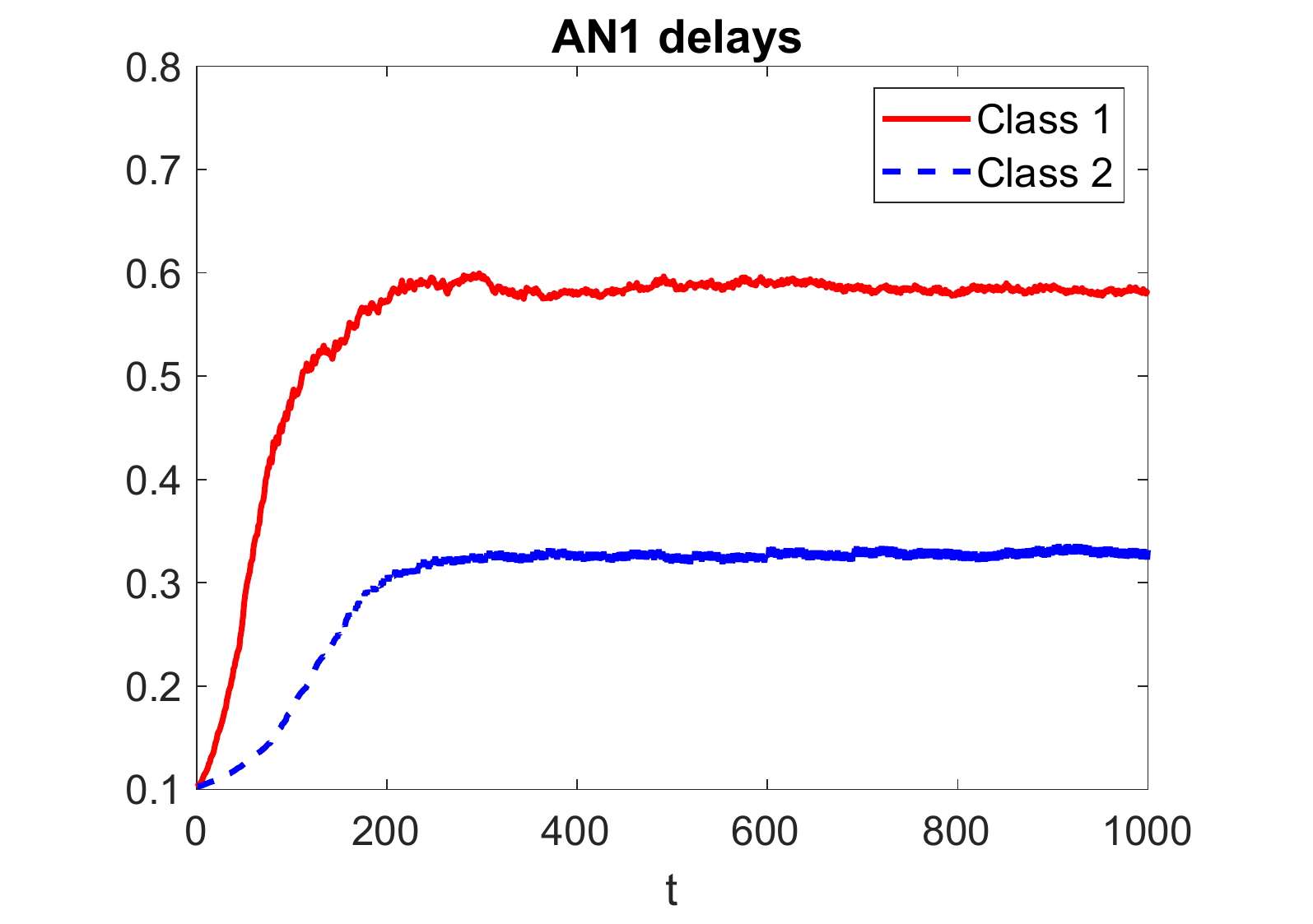}\hspace{-7mm}
\includegraphics[width=2in]{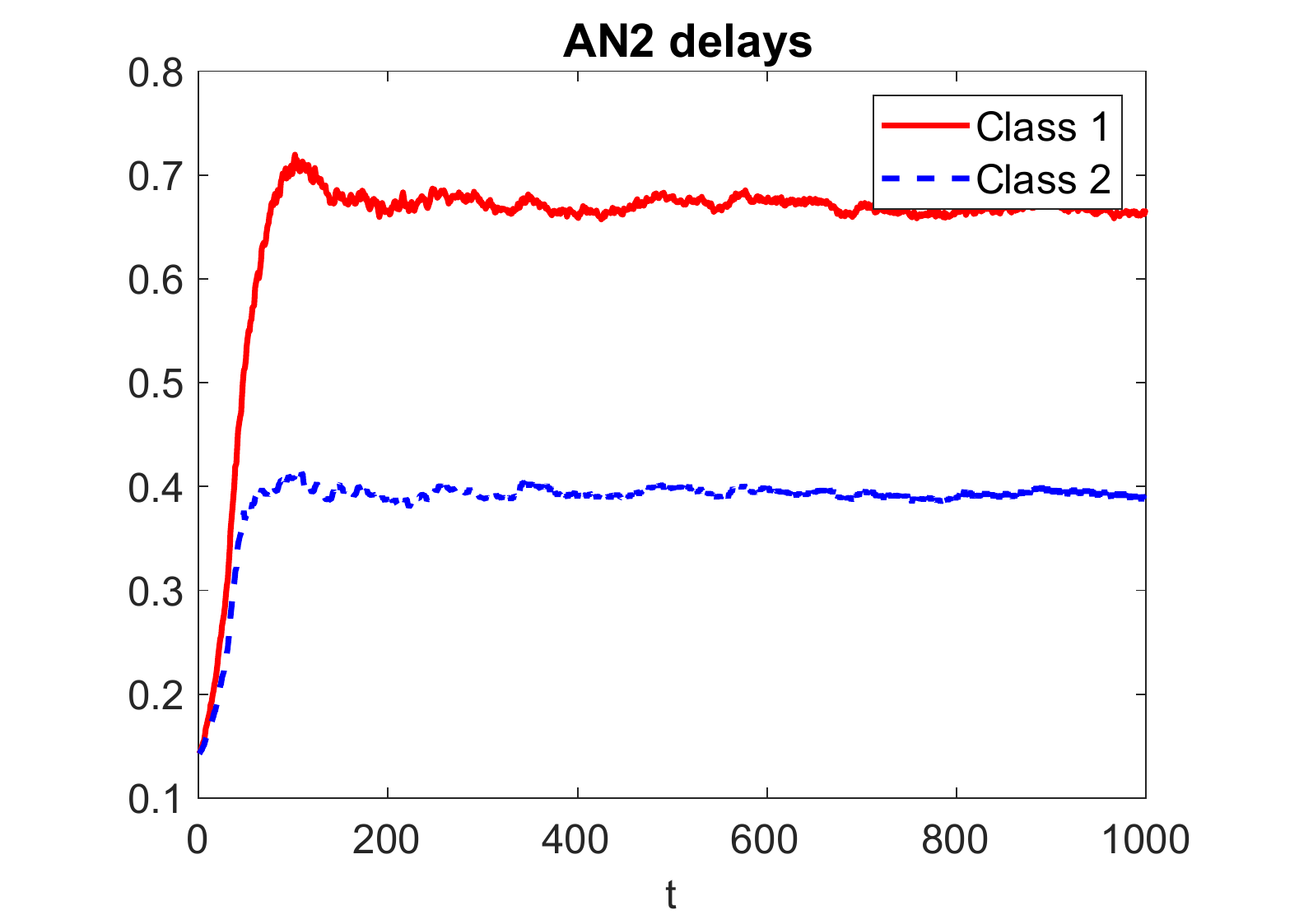}
}
\caption{End-to-end delays experienced by
traffic traversing to AN 1 (left) and AN 2
(right).}
\label{fig:Delay}
\end{figure}

Fig.~\ref{fig:Delay} plots the realized E2E delays 
experienced by the traffic destined for AN 1
(left) and AN 2 (right). With the employed parameters, the true delay constraints are satisfied for both traffic classes. In contrast, when there is no penalty function, the realized E2E delays for class 1 and 2 traffic destined for AN 1 (resp. AN 2) are 3.7 and 8.7 
(resp. 5.6 and 10.6), respectively. Obviously, these numbers are many times larger than their delay budgets and, hence, are unacceptable. 

\begin{figure}[h]
\centerline{
\includegraphics[width=2in]{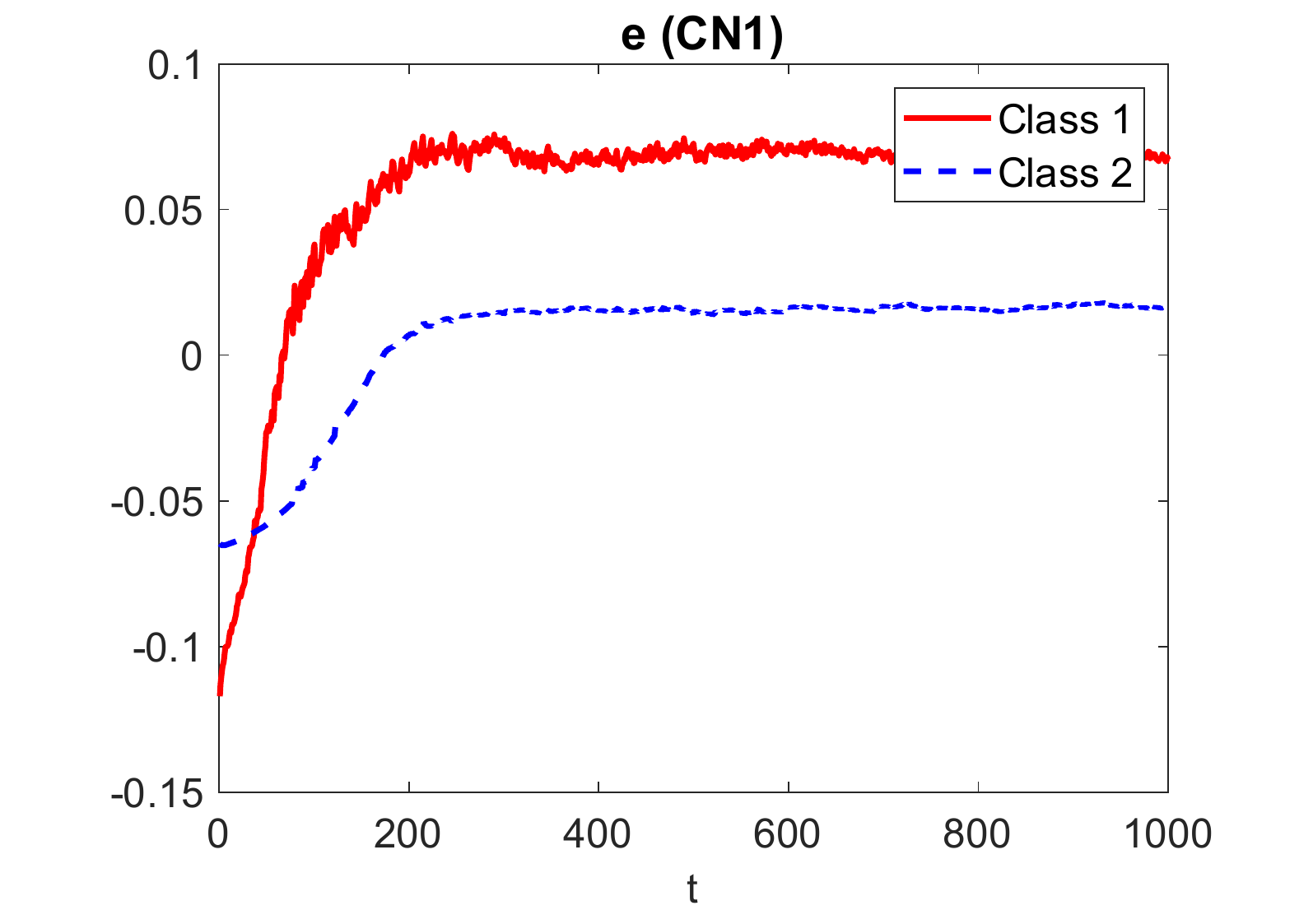}\hspace{-7mm}
\includegraphics[width=2in]{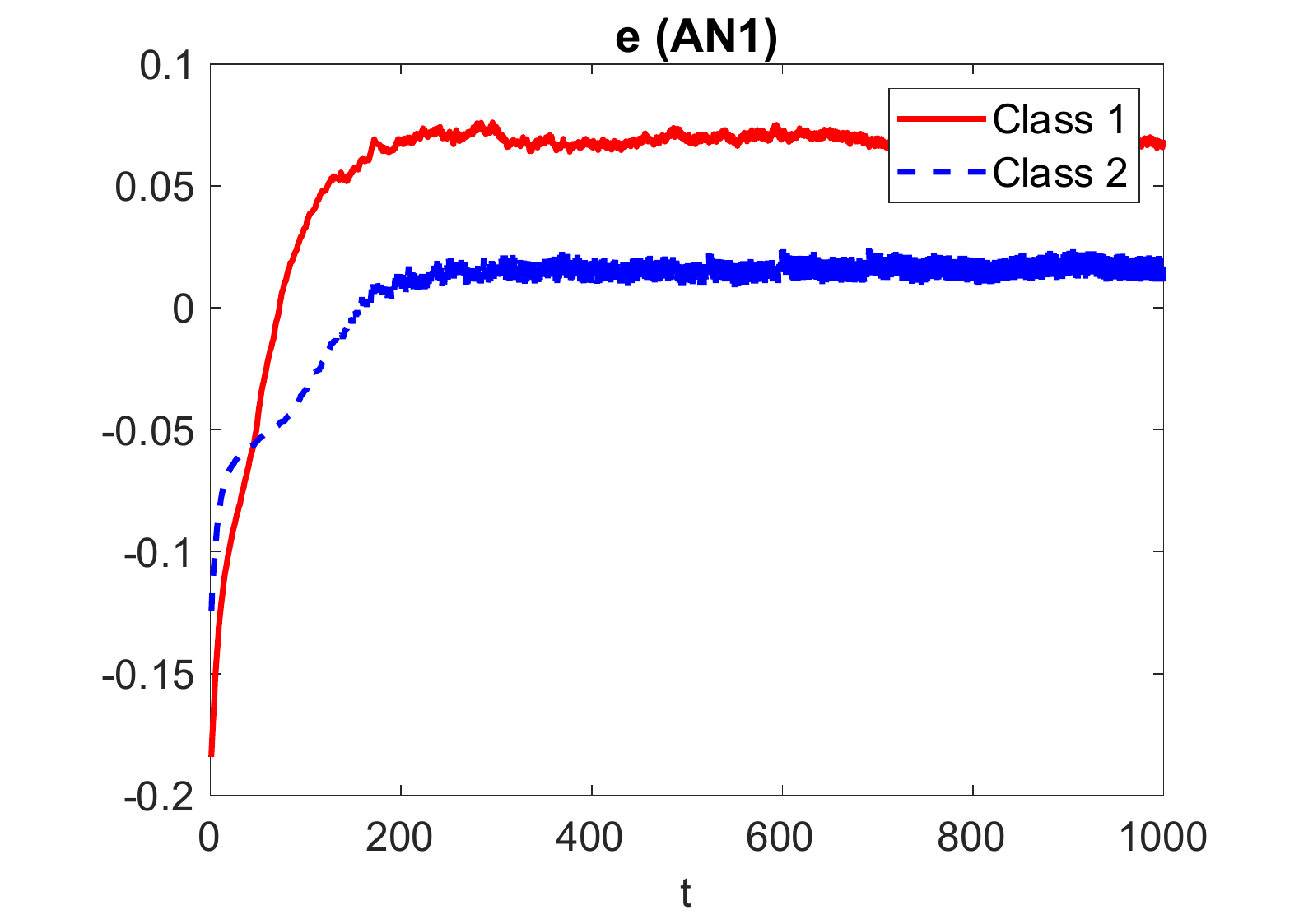}
}
\centerline{
\includegraphics[width=2in]{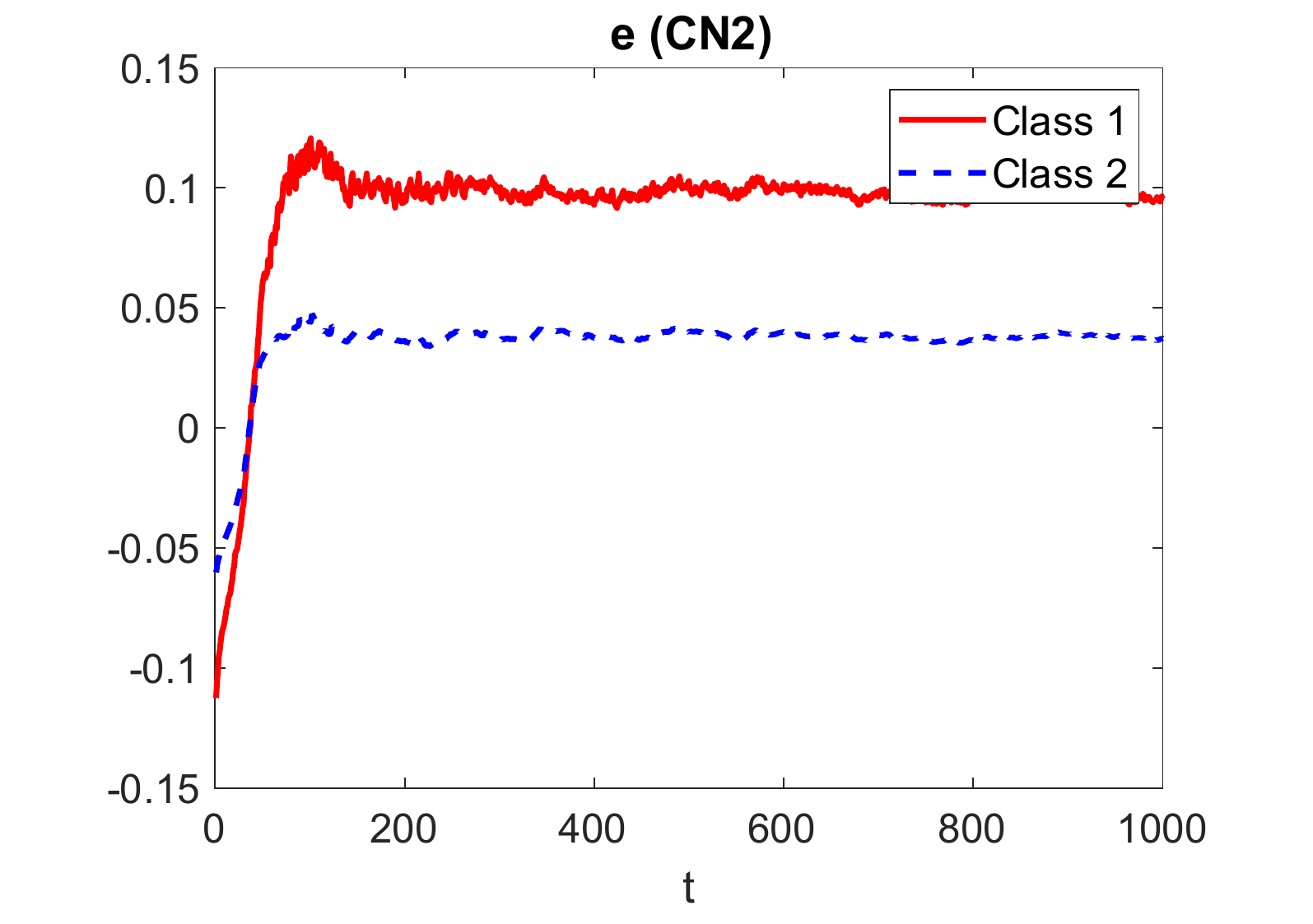}\hspace{-7mm}
\includegraphics[width=2in]{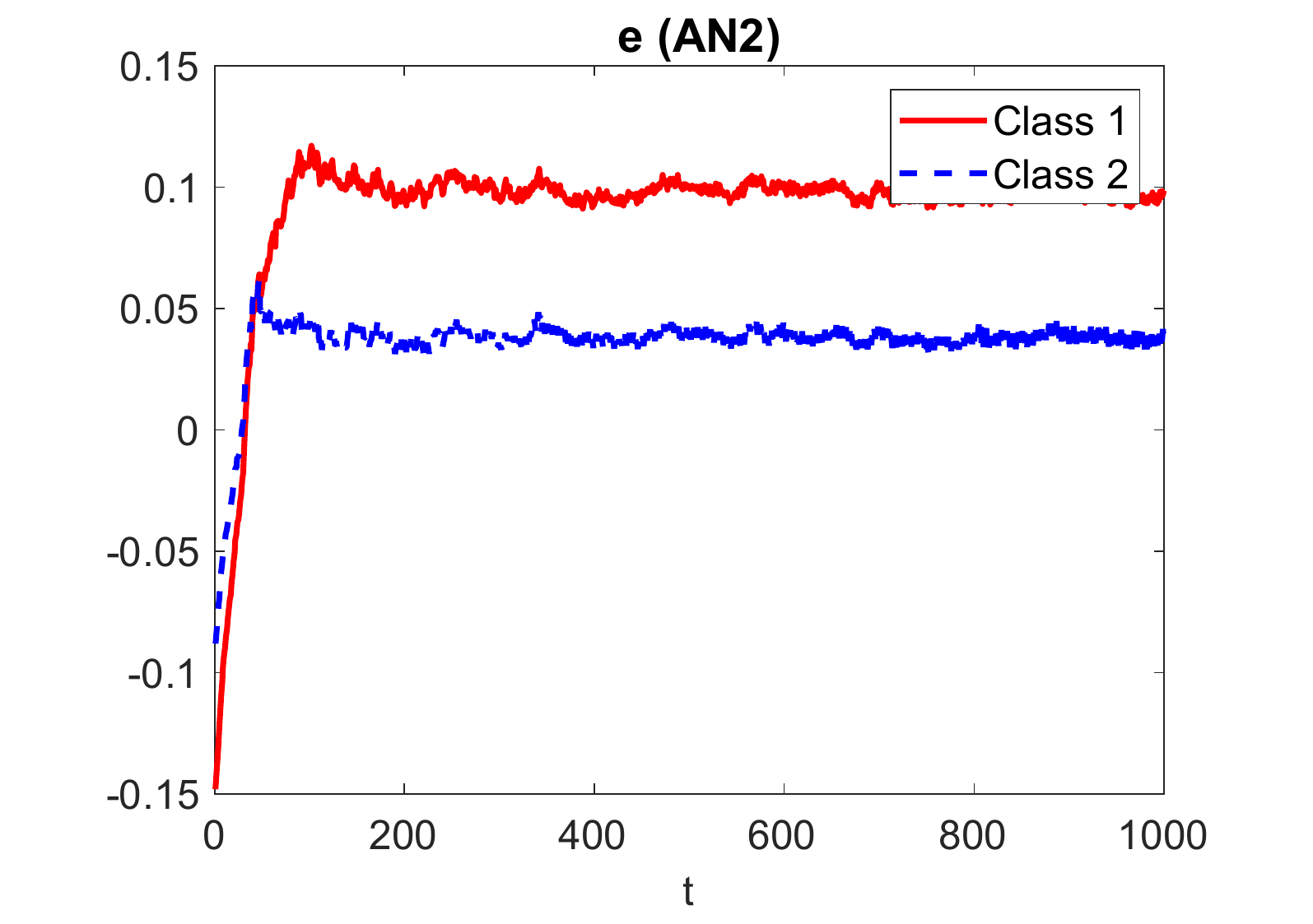}
}
\caption{Estimates of constraint functions
maintained by agents.}
\label{fig:Estimate}
\end{figure}

Recall that the E2E target delay budgets for class 
1 and 2 traffic are 
$D^T_1 = 0.42$ and $D^T_2 = 0.3$, respectively. 
The delays experienced by the traffic destined for 
AN 1 in the left
plot of Fig.~\ref{fig:Delay}, which include the
delays in both the CN and AN 1, exceed the 
target delays by approximately 0.17 and 0.03 for
class 1 and 2 traffic, respectively. 
Hence, according to~\eqref{eqAverageError}, 
the estimates of the agents must be approximately
0.06 and 0.01. Similarly, according 
to the right plot in Fig.~\ref{fig:Delay}, 
the agents' estimates for class 1 and 2 traffic
should be approximately 0.1 and 0.03. 
Fig.~\ref{fig:Estimate} shows the estimates 
$\eb_i(t)$, $i \in \{c, 1, 2\}$; the top figures 
(resp. bottom figures) are the estimates of
$g_{1, s}$ (resp. $g_{2, s}$), $s = 1, 2$. 
It is clear from the figure that the agents 
are indeed able to closely track the average 
constraint function values using the proposed 
consensus-type algorithm in \eqref{eq:update_e}.

\section{Discussion: Relation to Nash Equilibria
    of A State-Based Potential Game}
    \label{sec:Game}

In the previous sections, we implicitly assumed that the agents are willing to employ the proposed algorithm to cooperate in order to satisfy global constraints. In general, the agents will adopt the proposed algorithm
only if they believe that they cannot reduce their own
costs without significantly increasing global constraint violations by deviating from the equilibrium, i.e., an 
optimal point of \eqref{eq:ApproximatedProblem}. It turns
out that there is a close relation between an optimal 
point of \eqref{eq:ApproximatedProblem} and a Nash 
equilibrium of a state-based potential game (SBPG) associated with the problem.

An SBPG consists of the following~\cite{marden2015game}: 
\bitem 
\item[a)] a set of agents $\cA$; 
\item[b)] a state space ${\cal X}$; 
\item[c)] for each agent $i \in \cA$ and each state $\bx \in 
{\cal X}$, (i) a state-dependent admissible action set 
${\cal Y}_i(\bx)$ and (ii) a state-dependent cost 
function $J_i$, where $J_i(\bx, {\bf a})$ is agent $i$'s 
cost at the state $\bx$ when the agents adopt action 
profile ${\bf a} \in \prod_{i \in \cA} {\cal Y}_i(\bx) =: 
{\cal Y}(\bx)$; and
\item[d)] a deterministic state
transition function $f$: given the current state $\bx$
and agents' action profile ${\bf a} \in {\cal Y}(\bx)$, 
the next state is given by $f(\bx, {\bf a}) \in {\cal X}$.
\eitem 
There is also a null action profile ${\bf a}^0$ such 
that ${\bf a}^0 \in {\cal Y}(\bx)$ and $\bx = f(\bx, {\bf a}^0)$ 
for all $\bx \in {\cal X}$. In other words, the state does not 
change when the agents adopt the null action profile. 
A state-action pair $(\bx^\star, {\bf a}^\star)$ is
said to be a {\em stationary state Nash equilibrium}
of the SBPG if (i) $a_i^\star \in \arg \min_{a_i \in 
{\cal Y}_i(\bx^\star)} J_i\big(\bx^\star, (a_i, {\bf a}_{-i}^\star) \big)$ for every agent $i \in \cA$, 
where ${\bf a}_{-i}^\star$ is the action profile
of all agents except for agent $i$, i.e., 
${\bf a}_{-i}^\star = (a_j^\star : j \in 
\cA \setminus \{i\})$, and (ii) 
$\bx^\star = f(\bx^\star, {\bf a}^\star)$.
 
Consider a game with the state given by $\bx = (\by, \be)$, where $\by$ is the vector of decision variables in our 
optimization problem, and $\be = (\be_i : i \in \cA)$ is
the vector of constraint function estimates. 
Given a state $\bx \in {\cal X}$, an action $a_i$ of agent $i$
is a tuple $\big( \hat{\by}_i, (\hat{\be}_{i \to j} : 
j \in {\cal N}_i) \big)$, where $\hat{\by}_i$ is the change
in agent $i$'s decision variables and $\hat{\be}_{i \to j}$
is the estimates of constraint functions agent $i$ forwards
to its neighbor $j \in \mathcal{N}_i := \{j \in \cA \ | \ 
w_{ij} > 0\}$. Suppose that the state transitions to a new 
state based on the current state $\bx$ and the chosen 
action profile ${\bf a}$ as follows:
\beqan
\overline{\by}_i 
\myeq \by_i + \hat{\by}_i  \\
\overline{\be}_i 
\myeq \be_i + \bg_i(\overline{\by}_i) - \bg_i(\by_i)
+ 
\textstyle \sum_{j \in {\cal N}_i} \big( \hat{\be}_{j \to i}
    - \hat{\be}_{i \to j} \big).
\eeqan
Consider the cost function of agent $i$ given by 
\beqa
J_i(\bx, \ab) 
= \phi_i(\overline{\yb}_{i}) + \frac{\mu}{2} \textstyle
    \sum_{j \in \mathcal{N}_i} \| [\overline{\eb}_j]_+\|_2^2, 
    \label{eqNodalCost_PF}
\eeqa 
where $\mu > 0$ is a trade-off parameter. This gives rise 
to an SBPG with a global potential function given by
\cite{LiMarden14} 
\begin{equation}
\Phi_{\mu}(\bx, \ab) 
= \textstyle \sum_{i \in \mathcal{A}} \phi_i(\overline{\yb}_{i}) 
    + \frac{\mu}{2} \textstyle \sum_{i \in \mathcal{A} } 
        \| [\overline{\eb}_i]_+ \|_2^2. 
	        \label{eqPotentialFunc}
\end{equation}

Note the similarity between the objective function in 
\eqref{eq:ApproximatedProblem} and the global 
potential function in \eqref{eqPotentialFunc}. From 
\eqref{eqAverageError}, we have $\bg(\by(t)) = 
\sum_{i \in \cA} \bg_{i}(\by_i(t)) = \sum_{i \in \cA} \be_i(t)$. 
Hence, when all agents have the correct estimate of the
average constraint function values, i.e., $\be_i(t) = 
\be_\avg(t)$ for all $i \in \cA$, 
the global potential function coincides with the 
objective function in \eqref{eq:ApproximatedProblem}
with a different penalty parameter. This observation
can be used to prove the following proposition.
\begin{prop}
Suppose that $\by^* \in {\cal Y}^*$ is an optimal 
point of \eqref{eq:ApproximatedProblem}. Then, 
$\big( (\by^*, \gb(\by^*)/N), \0 \big)$, where
$\0$ is the null action profile, 
is a stationary state Nash equilibrium of the SBPG. 
\end{prop}

The above proposition means that each optimal point
in ${\cal Y}^*$ leads to a stationary state Nash 
equilibrium and 
there is an one-to-one mapping from ${\cal Y}^*$ 
to a set of stationary state Nash equilibria of the 
SBPG.

\section{Conclusions}
    \label{sec:Conclusion}
    
We proposed a novel distributed algorithm that enables autonomously managed ANs and CNs in a telecommunication system to cooperate with each other to meet E2E KPI goals. The new algorithm overcomes two major limitations of prior techniques. First, unlike prior approaches that require static  local KPI budgets  for  each autonomous  subsystem,  the  new  algorithm  allows the  autonomous  subsystems  to  dynamically negotiate  their  local  KPI  budgets.  Second, rather than requiring the autonomous subsystems to share their local decision variables with each other, our proposed  algorithm  only  needs  the  subsystems  to  exchange  their estimates of the global constraint functions. We proved that  the  new  algorithm  converges  to  an  optimal solution  almost surely, and presented  numerical  results  to  demonstrate that  the  convergence occurs quickly even with measurement noise.


\begin{thebibliography}{99}


\bibitem{TS22.261}
3GPP TS 22.261 V15.8.0 (2019-09), ``3rd Generation Partnership Project; Technical Specification Group Services and System Aspects; Service requirements for the 5G system; Stage 1 (Release 15).”

\bibitem{TS23.501}
3GPP TS 23.501 V15.12.0 (2020-12), ``3rd Generation Partnership Project; Technical Specification Group Services and System Aspects; System architecture for the 5G System (5GS); Stage 2 (Release 15)."


\bibitem{TS28.550}
3GPP TS 28.550 V16.1.0 “Management and orchestration; Performance assurance”.

\bibitem{TS28.552}
3GPP TS 28.552 V16.1.0 “Management and orchestration; 5G performance measurements”.

\bibitem{TS28.554}
3GPP TS 28.554 V16.0.0 “Management and orchestration; 5G end to end Key Performance Indicators (KPI)”.


\bibitem{TS32.425}
3GPP TS 32.425 V16.3.0 “Performance Management (PM); Performance measurements for Evolved Universal Terrestrial Radio Access Network (E-UTRAN)”.

\bibitem{TS28.532}
3GPP TS 28.532 V16.0.0 “Management and orchestration; Generic management services”.



\bibitem{Asad20}
M. Asad, A. Basit, S. Qaisar and M. Ali, ``Beyond 5G: Hybrid End-to-End Quality of Service Provisioning in Heterogeneous IoT Networks," {\em IEEE Access}, 8:192320-192338, 2020, doi: 10.1109/ACCESS.2020.3032704.


\bibitem{Bianchi13}
P. Bianchi and J. Jakubowicz, 
``Convergence of a multi-agent projected stochastic
    gradient algorithm for non-convex optimization,"
{\em IEEE Trans. on Automatic Control}, 58(2):391-405,


\bibitem{Boyd}
S. Boyd and L. Vandenberghe, 
{\em Convex Optimization}, 
Cambridge University Press, 2004.
    	



\bibitem{Dutra17}
D. L. C. Dutra, M. Bagaa, T. Taleb and K. Samdanis, ``Ensuring End-to-End QoS Based on Multi-Paths Routing Using SDN Technology," Proc. of IEEE GLOBECOM, Singapore, 2017, pp. 1-6, doi: 10.1109/GLOCOM.2017.8254076.


\bibitem{Jacquet2014}
R. Jacquet, G. Texier and A. Blanc, ``Computing end-to-end QoS paths in the Internet considering multiple alliances," Proc. of the 16th International Telecommunications Network Strategy and Planning Symposium (Networks), 2014, pp. 1-6, doi: 10.1109/NETWKS.2014.6959248.

\bibitem{KushYin}
H. Kushner and G. Yin,
{\em Stochastic Approximation and Recursive Algorithms and Applications}, 2nd ed. 
Springer, 2003. 

\bibitem{Lee19}
Y. Lee, K. John and R. Vilalta, ``Extended ACTN Architecture to Enable End-To-End 5G Transport Service Assurance," Proc. of the 21st International Conference on Transparent Optical Networks (ICTON), Angers, France, 2019, pp. 1-3, doi: 10.1109/ICTON.2019.8840270.

\bibitem{Letaief19}
K. B. Letaief, W. Chen, Y. Shi, J. Zhang and Y. A. Zhang, ``The Roadmap to 6G: AI Empowered Wireless Networks," {\em IEEE Communications Magazine}, 57(8):84-90, Aug. 2019, doi: 10.1109/MCOM.2019.1900271.

\bibitem{LiMarden14}
N. Li and J.R. Marden, 
``Decoupling coupled constraints through utility design,"
{\em IEEE Trans. on Automatic Control}, 59(8):2289-=2294, 
Aug. 2014. 

\bibitem{Mai2021}
V. S. Mai, R. J. La, T. Zhang and A. Battou, ``Distributed Optimization with Global Constraints Using Noisy Measurements," arXiv preprint arXiv:2106.07703, 2021.

\bibitem{marden2015game}
J. Marden and J. S. Shamma, ``Game theory and distributed control," {\em Handbook of Game Theory with Economic Applications}, pp. 861-899, 2015. 

\bibitem{Saad20}
W. Saad, M. Bennis and M. Chen, ``A Vision of 6G Wireless Systems: Applications, Trends, Technologies, and Open Research Problems," {\em IEEE Network}, vol. 34(3):134-142, May/June 2020, doi: 10.1109/MNET.001.1900287.


\bibitem{She20}
C. She et al., ``Deep Learning for Ultra-Reliable and Low-Latency Communications in 6G Networks," {\em IEEE Network}, 34(5):219-225, Sep./Oct. 2020, doi: 10.1109/MNET.011.1900630.



\bibitem{Srivastava11}
K. Srivastava and A. Nedi$\acute{\rm c}$, 
``Distributed asynchronous constrained stochastic
    optimization,"
{\em IEEE Journal of Selected Topics in Signal Processing}, 5(4:)772-790, 2011.


\bibitem{Teng20}
C.-C. Teng, M.-C. Chen, M.-H. Hung and H.-J. Chen, ``End-to-end Service Assurance in 5G Crosshaul Networks," Proc. of the 21st Asia-Pacific Network Operations and Management Symposium (APNOMS), Daegu, Korea (South), 2020, pp. 306-309, doi: 10.23919/APNOMS50412.2020.9236977.

\bibitem{Ye18b}
Q. Ye, J. Li, K. Qu, W. Zhuang, X. S. Shen and X. Li, ``End-to-End Quality of Service in 5G Networks: Examining the Effectiveness of a Network Slicing Framework," {\em IEEE Vehicular Technology Magazine}, 13(2):65-74, Jun. 2018, doi: 10.1109/MVT.2018.2809473.

\bibitem{Zhang2020}
T. Zhang, ``Toward Automated Vehicle Teleoperation: Vision, Opportunities, and Challenges," {\em IEEE Internet of Things Journal}, 7(12):11347-11354, Dec. 2020, doi: 10.1109/JIOT.2020.3028766.

\bibitem{Zhu19}
G. Zhu, J. Zan, Y. Yang and X. Qi, ``A Supervised Learning Based QoS Assurance Architecture for 5G Networks," {\em IEEE Access}, 7:43598-43606, 2019, doi: 10.1109/ACCESS.2019.2907142.


\end{thebibliography}
\end{document}